\documentclass[twocolumn,english,aps,prl,showpacs]{revtex4}
\usepackage[T1]{fontenc}
\usepackage[latin1]{inputenc}
\usepackage{amsmath}
\usepackage{graphicx}
\usepackage{amssymb}
\usepackage{esint}



\usepackage{babel}

\begin{document}

\title{Cavity-Assisted Dynamical Spin-Orbit Coupling in Cold Atoms}

\author{Lin Dong$^{1}$, Lu Zhou$^{1,2}$, Biao Wu$^{3}$, B. Ramachandhran$^{1}$ and Han Pu$^{1}$}

\affiliation{$^{1}$Department of Physics and Astronomy, and Rice Quantum Institute,
Rice University, Houston, TX 77251, USA \\
 $^{1,2}$Department of Physics, and Quantum Institute for Light and Atoms, East China Normal University, Shanghai 200062, China\\
 $^{3}$International Center for Quantum Materials, Peking University, Beijing 100871, China }

\date{\today}
\begin{abstract}
We consider ultracold atoms subjected to a cavity-assisted two-photon Raman transition. The Raman coupling gives rise to effective spin-orbit interaction which couples the atom's center-of-mass motion to its pseudospin degrees of freedom. Meanwhile, the cavity photon field is dynamically affected by the atom. This feedback between the atom and photons leads to a dramatic modification of the atomic dispersion relation, and further leads to dynamical instability of the system. We propose to detect the change in the cavity photon number as a direct way to demonstrate dynamical instability.

\end{abstract}


\maketitle

\emph{Introduction} --- When an atom interacts with a quantized light field supported by an optical cavity, the atom and the light field mutually affect each other. A self-consistent solution for the light field and the atom is thus required. This has been a major theme in cavity quantum electrodynamics (CQED) \cite{CQED}. In traditional CQED settings, only the internal dynamics of the atom is relevant. In recent years, ultracold atoms have been put inside optical cavities and in such a situation, one can no longer neglect the center-of-mass (COM) motion of the atom. A variety of phenomena in this ``ultracold atom + cavity" system, which is  an example of an optomechanical system, has been explored experimentally \cite{cv1,cv2,cv3,zimm,cv4,collect2,dan1} and theoretically \cite{collect1,review}.  

\begin{figure}[b]
\includegraphics[width=.48\textwidth]{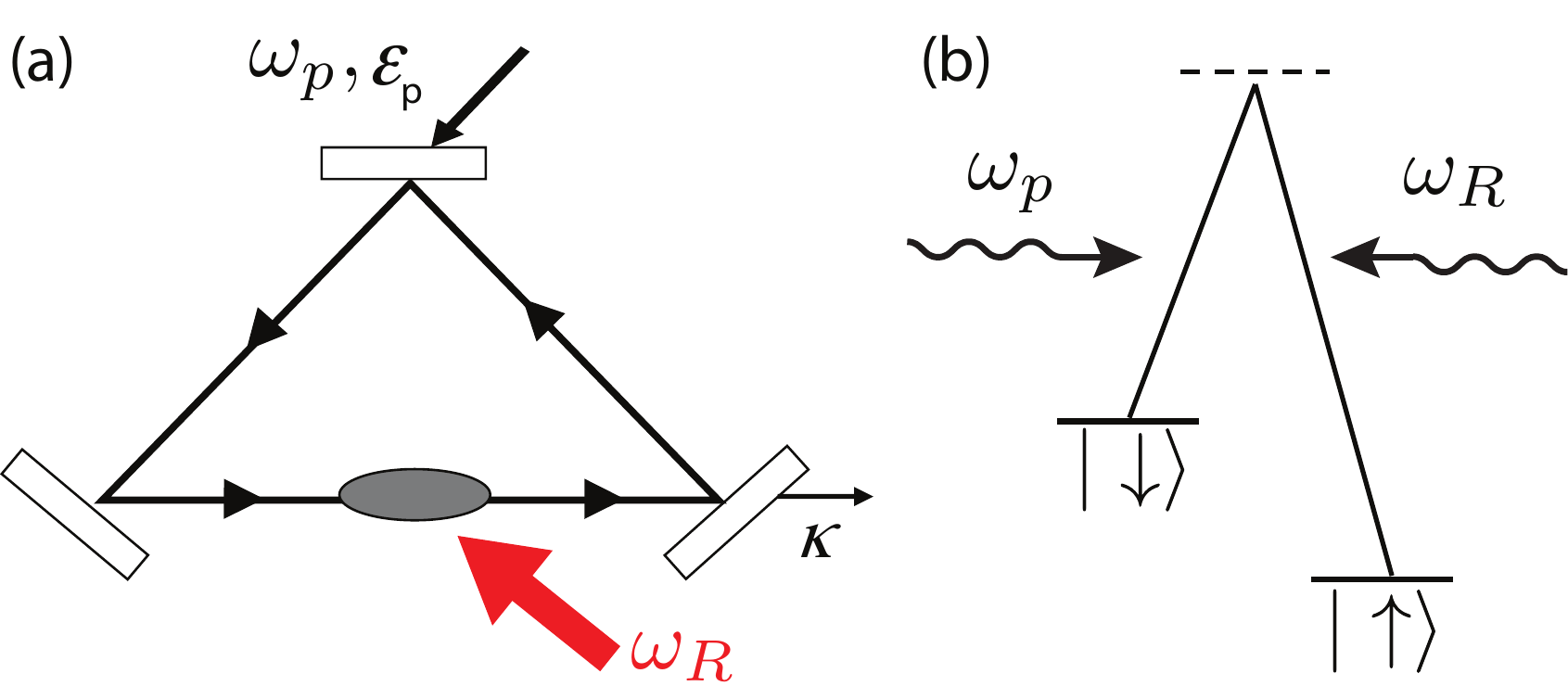}\caption{(Color Online) (a) Schematic diagram of the cavity-assisted spin-orbit coupled system; (b) Level diagram of atom photon/light field interaction.}\label{scheme}
\end{figure}

Another recent breakthrough in cold atom research is the realization of spin-orbit coupling (SOC) in ultracold atoms, in both bosonic \cite{lin} and fermionic systems \cite{Zhang, mit}, (refer to \cite{soc_reviews} for reviews). Realization of SOC in cold atoms involves a two-photon Raman transition between two hyperfine ground states as schematically shown in Fig.~\ref{scheme}(b). The Raman-induced transition between the two atomic levels and the associated momentum transfer due to photon recoil give rise to an effective coupling between the COM motion and the internal states of the atom. This SOC underlies numerous novel phenomena, ranging from spin Hall effects to topological insulators. 

In the experiments of spin-orbit coupled quantum gases, the Raman beams that generate the SOC are provided by two classical laser fields, which are not affected by the atoms. Here we consider a situation where one of the Raman beams is replaced by a quantized light field supported by an optical cavity, as schematically shown in Fig.~\ref{scheme}(a). In this scheme, akin to other ``ultracold atom + cavity" systems, there will be a back-action from the atom to the light. Therefore the SOC in the atom is generated by a quantized light field which itself is affected by the atomic dynamics. In this sense, the cavity-assisted SOC becomes {\em dynamic} \cite{note,gaugeTheory}. Furthermore, in previous experimental studies of cold-atom-based cavity optomechanical systems \cite{cv1,cv2,cv3,zimm,cv4,collect2,dan1}, only the COM motion of the atom is included. The inclusion of the internal spin degrees of freedom and the resulting SOC opens up a new avenue of research in cavity optomechanical systems. We will show that this dynamic SOC dramatically modifies the atomic dispersion relation and the stability of the system.     


\emph{Model and formalism} --- We consider a single atom (or, a non-interacting Bose-Einstein condensate) with two relevant internal states (denoted as $|\uparrow \rangle$ and $|\downarrow \rangle$) confined inside a unidirectional optical ring cavity, depicted schematically in Fig.~\ref{scheme}. The cavity is pumped by a coherent laser field with frequency $\omega_p$ and pumping rate $\varepsilon_p$. It supports a single mode traveling wave and has an intrinsic angular frequency $\omega_c$. An additional coherent laser beam with frequency $\omega_R$ shines on the atom, which together with the cavity field provides the Raman transition between $|\uparrow \rangle$ and $|\downarrow \rangle$ states. During the Raman transition, a recoil momentum of $\pm 2\hbar q_r \hat{z}$ is transferred to the atom. We treat the leakage of cavity photon phenomenologically by introducing a cavity decay rate $\kappa$. The model Hamiltonian is thus written as (we take $\hbar=1$),
\begin{eqnarray}
H & = & \sum_{\sigma}\int d{\bf r}\left[\Psi_{\sigma}^{\dagger}({\bf r})\left(\frac{\hat{{\bf k}}^{2}}{2m}+\epsilon_{\sigma}^{0}\right)\Psi_{\sigma}({\bf r})\right]\nonumber \\
 & + & \frac{\Omega}{2}\int d{\bf r} \, e^{+2i q_{r}z}\Psi_{\uparrow}^{\dagger}({\bf r})\Psi_{\downarrow}({\bf r})\tilde{c}e^{+i\omega_{R}t}\nonumber \\
 & + & \frac{\Omega}{2}\int d{\bf r}\,e^{-2i q_{r}z} \tilde{c}^{\dagger} \Psi_{\downarrow}^{\dagger}({\bf r})\Psi_{\uparrow}({\bf r}) e^{-i\omega_{R}t}\nonumber \\
 & + & i\varepsilon_{p}(\tilde{c}^{\dagger}e^{-i\omega_{p}t}-\tilde{c}e^{+i\omega_{p}t})+\omega_{c}\tilde{c}^{\dagger}\tilde{c}-i\kappa \tilde{c}^\dagger \tilde{c}, \label{bareH}
 \end{eqnarray}
where $\Psi_\sigma({\bf r})$ ($\sigma = \uparrow$, $\downarrow$) is the atomic annihilation  operator, $\epsilon_\sigma^0$ is the corresponding bare atomic energy, and $\tilde{c}$ represents the photon annihilation operator. $\Omega$ describes the atom-photon coupling strength. However, the true Raman coupling strength also includes the cavity photon amplitude of $\tilde{c}$ or $\tilde{c}^\dag$ which is coupled to the atomic operators. It is this coupling that renders the resulting SOC \emph{dynamic}.
 
 It is convenient to work work in a frame rotating at pump laser frequency $\omega_p$ by transforming the photon operator to $c=\tilde{c}e^{i\omega_p t}$. This is equivalent to performing an unitary transformation $U=e^{+i\omega_p\tilde{c}^\dagger \tilde{c}t}$ to the Hamiltonian~(\ref{bareH}), by $H'=UHU^{-1}+i\frac{dU}{dt}U^{-1}$. From $H'$, we perform another unitary transformation $\tilde{U}=e^{i\delta_Rt(\Psi^\dagger_\uparrow\Psi_\uparrow-\Psi^\dagger_\downarrow\Psi_\downarrow)/2}$, with $\delta_R=\omega_p-\omega_R$, to obtain the Hamiltonian $H''$. Finally, after a gauge transformation to atomic operators $\psi_\uparrow=\Psi_\uparrow e^{-iq_rz}$ and $\psi_\downarrow=\Psi_\downarrow e^{+iq_rz}$, we arrive at the following effective Hamiltonian $\mathcal{H}_{\rm eff}$: 
 \begin{eqnarray}
 \mathcal{H}_{\rm eff}& = & \sum_\sigma\int d{\bf r}\left[ \psi^\dagger_\sigma({\bf r})\left(\frac{ \hat{{\bf k}}^2+ 2\alpha q_r {k}_z}{2m}+\alpha\tilde{\delta}\right)\psi_\sigma({\bf r})\right]\nonumber \\
 & + & \frac{\Omega}{2}\int d{\bf r}\left[{\psi}_{\uparrow}^{\dagger}({\bf r}){\psi}_{\downarrow}({\bf r})c+c^{\dagger}{\psi}_{\downarrow}^{\dagger}({\bf r}){\psi}_{\uparrow}({\bf r}) \right]\nonumber \\
 & + & i\varepsilon_{p}(c^{\dagger}-c)-\delta_c c^{\dagger}c-i\kappa c^{\dagger}c, \label{effH}
 \end{eqnarray}
where $\tilde{\delta}=\delta_R/2+(\epsilon_\uparrow^0-\epsilon_\downarrow^0)$ represents the two-photon Raman detuning; $\delta_c=\omega_p-\omega_c$ is the cavity-pump detuning, and $\alpha=\pm 1$ for $\sigma=\uparrow,\downarrow$, respectively.  


\emph{Dispersion Relation.} --- From the Hamiltonian~(\ref{effH}), one can easily obtain the following equations of motion (EOM): 
\begin{eqnarray}
i\frac{d}{dt}c 
 & = & i\varepsilon_{p}-(\delta_{c}+i\kappa)c+\frac{\Omega}{2}\int d{\bf r}\psi_{\downarrow}^{\dagger}({\bf r})\psi_{\uparrow}({\bf r}), \label{EOMc}\\
i\frac{\partial}{\partial t} \psi({\bf r}) 
 & = & \left(\begin{array}{cc}
\frac{\hat{{\bf k}}^2+2q_{r} {k}_{z}}{2m}+\tilde{\delta} & \frac{\Omega}{2}c\\
\frac{\Omega}{2}c^{\dagger} & \frac{\hat{{\bf k}}^2-2q_{r} {k}_{z}}{2m}-\tilde{\delta}\end{array}\right)\psi({\bf r})\,,\label{EOMpsi}
\end{eqnarray}
where $\psi({\bf r})\equiv[\psi_\uparrow({\bf r}), \psi_\downarrow({\bf r})]^T$.
To proceed further, we adopt a mean-field approximation by replacing the operators by their respective expectation values:
$c \rightarrow \langle c \rangle \,,\;\;\; \psi_\sigma ({\bf r}) \rightarrow \langle \psi_\sigma ({\bf r}) \rangle \equiv \varphi_\sigma ({\bf r})$,
which is valid for small quantum fluctuations of both operators $c$ and $\psi_\sigma({\bf r})$. Assuming a homogeneous atomic density distribution, we take 
the plane-wave ansatz for the atomic modes $\varphi_\sigma({\bf r})=e^{i{\bf k}\cdot{\bf r}}\varphi_\sigma$ with the normalization condition $|\varphi_\uparrow|^2+|\varphi_\downarrow|^2=1$. The steady-state solution for the photon field is given by  
\begin{equation}
\langle c\rangle=\frac{\varepsilon_{p}-\frac{i}{2}\Omega \varphi_\downarrow^* \varphi_\uparrow }{\kappa-i\delta_{c}}\,.\label{meanfieldc}
\end{equation}
Substituting Eq.~(\ref{meanfieldc}) into Eq.~(\ref{EOMpsi}), we have
\begin{eqnarray}
i\dot{\varphi}_{\uparrow}&\!\! =\!\! & \left(\frac{{\bf k}^{2}}{2m}+q_{r}k_{z}+\tilde{\delta}\right) \varphi_{\uparrow}+\frac{\Omega}{2}\frac{\varepsilon_{p}-\frac{i\Omega}{2} \varphi_\downarrow^\ast\varphi_\uparrow}{\kappa-i\delta_{c}}\varphi_{\downarrow} \,,\label{EOMphi1}\\
i\dot{\varphi}_{\downarrow} & \!\!= \!\!& \left(\frac{{\bf k}^{2}}{2m}-q_{r}k_{z}-\tilde{\delta}\right)\varphi_{\downarrow}+\frac{\Omega}{2}\frac{\varepsilon_{p}+\frac{i\Omega}{2} \varphi_\uparrow^\ast\varphi_\downarrow}{\kappa+i\delta_{c}}\varphi_{\uparrow}\,.\label{EOMphi2}
\end{eqnarray}

\begin{figure}[htp]
\includegraphics[width=.45\textwidth]{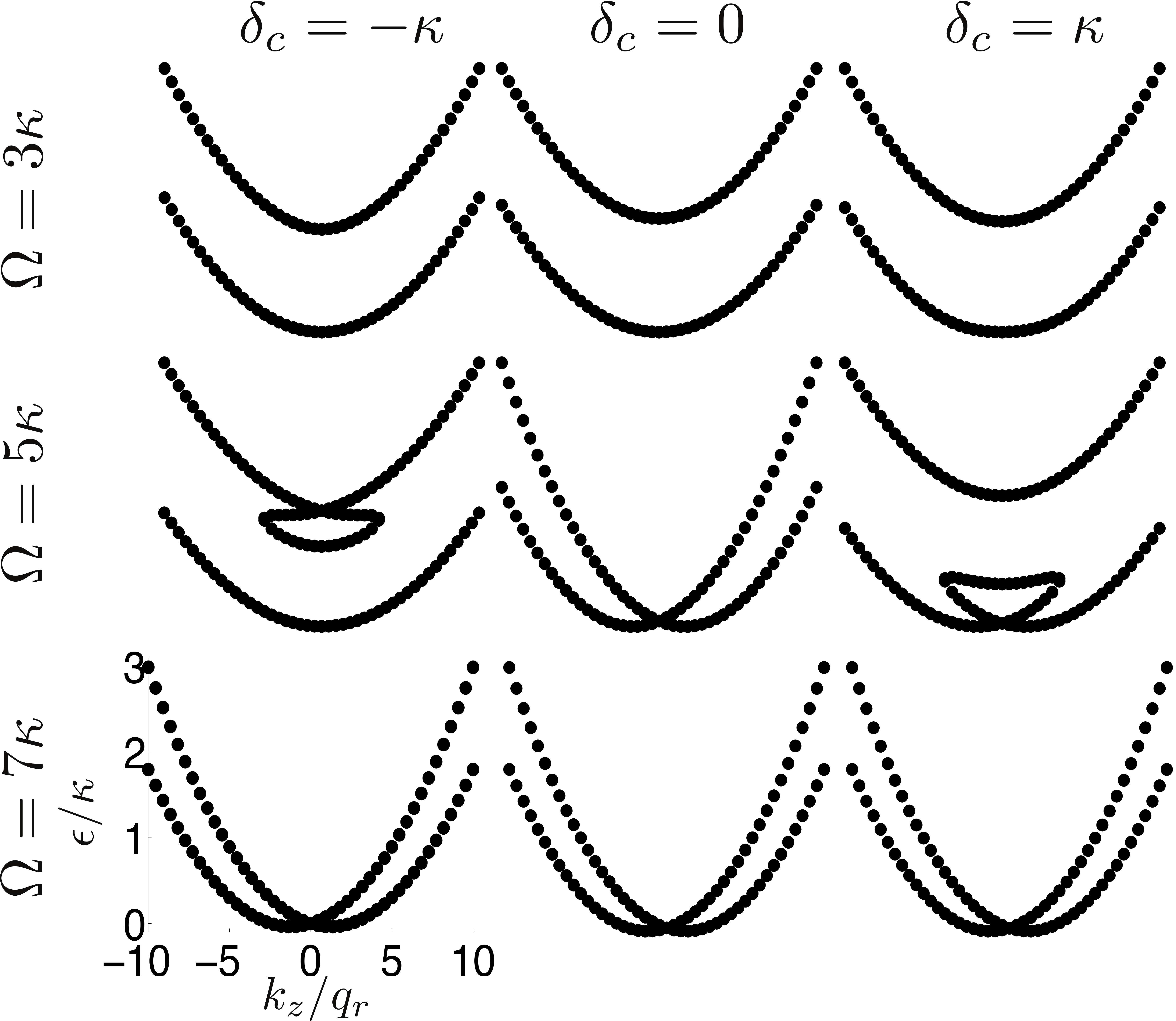}\caption{Eigenenergy $\epsilon$ as a function of quasi-momentum. We set $\tilde{\delta}=0$ and $\varepsilon_p=\kappa$. For nonzero $\delta_c$, a loop structure forms when $\Omega_c^{(1)} < \Omega < \Omega_c ^{(2)}$. For $\delta_c=\pm\kappa$, $\Omega_c^{(1)}=4\varepsilon_p$ and $\Omega_c^{(2)}= 4\sqrt{2}\varepsilon_p$. Throughout our calcultion, we take $\kappa$ and $\sqrt{2m\kappa}$ to be the units for energy and momentum, respectively. A typical value for $\kappa$ is $2\pi \times 1$ MHz, and we choose $q_r=0.22$ in our units. }\label{atomloops}
\end{figure}

For a given atomic quasi-momentum ${\bf k}$, we define energy levels as the solution of the time-independent version of Eqs.~(\ref{EOMphi1}) and (\ref{EOMphi2}), obtained by replacing $i(\partial/\partial t)$ with the eigenenergy $\epsilon({\bf k})$. After some calculation, we find that $\epsilon({\bf k})$ obeys a quartic equation: 
\begin{equation}
4\epsilon^4+B\epsilon^3+C\epsilon^2+D\epsilon+E=0 \,,
\label{generalquarticEq}
\end{equation}
where the detailed derivation and the expressions of coefficients are given in the Supplementary Material \cite{sup}.  
This quartic equation can be solved analytically, but the expressions are cumbersome. We plot the typical behavior of the dispersion relation $\epsilon({ k_z})$ vs $k_z$ for $\tilde{\delta}=0$ in Fig.~\ref{atomloops}. Note that we always take $k_x =k_y=0$, as the SOC only occurs along the $z$-axis. A maximum of four real roots are allowed by Eq.~(\ref{generalquarticEq}). As we will show, in such regimes, a loop structure develops in the dispersion curve.

As shown in Fig.~\ref{atomloops}, for $\delta_c=0$ (i.e., the pump field is resonant with the cavity), we always have two dispersion branches. The two branches are gapped when the atom-photon coupling strength $\Omega$ is small and touch each other at $k_z=0$ when $\Omega$ exceeds a critical value. For $\delta_c \neq 0$, we again have two gapped branches at small $\Omega$. As $\Omega$ is increased beyond a critical value, a loop appears near $k_z=0$ in either the upper or the lower branch depending on the sign of $\delta_c$. The loop increases in size as $\Omega$ increases and finally touches the other branch and dissolves when $\Omega$ reaches a second critical value. Note that such a dispersion relation is markedly different from that without the cavity, in which case one always obtains two gapped branches. The dispersion curves for finite $\tilde{\delta}$ are qualitatively similar, but in that case the curves are no longer symmetric about $k_z=0$ and the loop emerges  at finite $k_z$ (see Fig.~\ref{stability} below). 

\begin{figure}[htp]
\includegraphics[width=.48\textwidth]{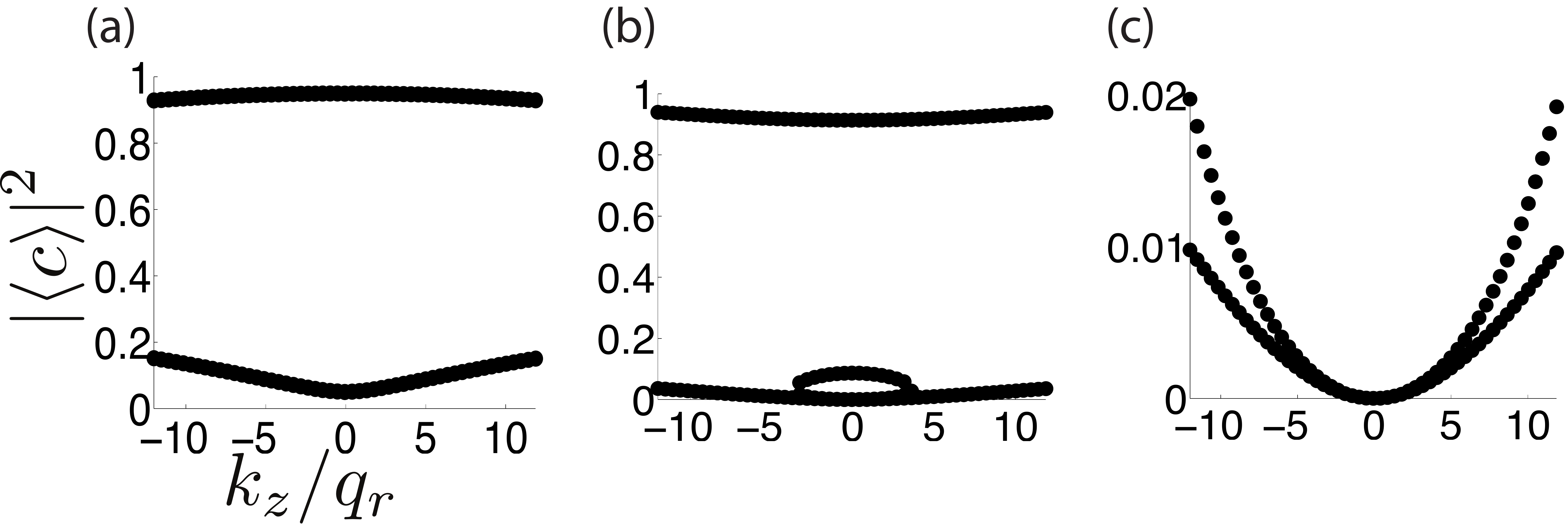}\caption{Photon number distribution as a function of atom's quasi-momentum. The parameters are the same as in the right column of Fig.~\ref{atomloops}, where $\delta_c=\kappa$, $\varepsilon_p=\kappa$, and   $\Omega/\kappa=3$, 5, and 7 from (a) to (c). }
\label{photonloops}
\end{figure}

The photon number distributions corresponding to the right column of Fig.~\ref{atomloops} are plotted in Fig.~\ref{photonloops}. As seen in Fig.~\ref{photonloops}(c), for sufficiently large $\Omega$, the cavity photon number decreases dramatically. Correspondingly, the effective Raman coupling becomes negligibly small, and the atomic dispersion curve becomes quadratic as in the absence of laser fields (see the bottom row of Fig.~\ref{atomloops}). This is analogous to the photon blockade phenomenon \cite{photonblock} in which the strong atom-photon coupling keeps pump photons from entering the cavity.

We can gain some insights about the general structure of the dispersion curve, and particularly the appearance and disappearance of the loop, by examining the quartic equation (\ref{generalquarticEq}) for $k_z=0$ and $\tilde{\delta}=0$. Under these conditions, Eq.~(\ref{generalquarticEq}) is simplified to:
\begin{equation}
\epsilon^2(4\epsilon^2-2w\epsilon+|v|^2-4|u|^2)=0\,,
\label{simplequartic}
\end{equation}
with the constraint that the root $\epsilon=0$ is only valid for $\Omega \ge 4\epsilon_p$ \cite{note1}.
Here the coefficients $w$, $u$ and $v$ are defined in \cite{sup}. Simple analysis shows that there are three regimes. First, 
when $\Omega < 4\varepsilon_p\equiv\Omega_c^{(1)}$, Eq.~(\ref{simplequartic}) has two real roots, one positive and one negative. This corresponds to the two gapped branches for small $\Omega$ in the top row of Fig.~\ref{atomloops}. Second, when $\Omega_c^{(1)} \leq \Omega \leq  4\varepsilon_p \sqrt{1+(\delta_c/\kappa)^2}\equiv\Omega_c^{(2)}$, Eq.~(\ref{simplequartic}) has four real roots --- two degenerate roots at $\epsilon=0$ and two additional roots with the same sign. This corresponds to the looped regime in the middle row of Fig.~\ref{atomloops}. Finally when $\Omega > \Omega_c^{(2)}$, only the two degenerate roots at $\epsilon=0$ exist, which correspond to the gapless regime represented by the bottom row in Fig.~\ref{atomloops}. Note that for $\delta_c=0$, we have $\Omega_c^{(1)}=\Omega_c^{(2)}=4\epsilon_p$, and the loop never develops.

The emergence of the loop structure is a distinctive nonlinear feature of the system. We remark that similar loop structures or the associated hysteretic phenomena have been found in other nonlinear systems \cite{sup}. The nonlinearity may originate from the mean-field density-density interaction \cite{loopPapers} or from the cavity-induced feedback between atoms and photons \cite{loop1}. The case studied here corresponds to the latter situation. However, in previous studies of ``ultracold atom + cavity" systems \cite{loop1}, the interaction between the cavity photons and atoms is dispersive, and so it does not induce SOC directly. As we will show below, the system studied here possesses very different dynamical and stability properties.   

{\em Stability and Dynamical Analysis} --- Nonlinear systems usually possess intriguing stability properties. To examine the stability of the eigenstates obtained above, we introduce conjugate variables $p=|\varphi_\downarrow|^2-|\varphi_\uparrow|^2$ and $\theta=\text{angle}(\varphi_\downarrow)-\text{angle}(\varphi_\uparrow)$, which correspond to the spin magnetization and the relative phase between the two atomic spin states. The EOM for $p$ and $\theta$ can be easily derived from Eqs.~(\ref{EOMphi1}) and (\ref{EOMphi2}) \cite{sup}, 
from which we can readily obtain the fixed points $(p^\ast,\theta^\ast)$ by setting $\dot{p}=\dot{\theta}=0$, which correspond to the eigenstates obtained earlier. To check the stability, we linearize the equations around the fixed points by taking $p=p^{\ast}+\delta p$, $\theta=\theta^{\ast}+
\delta\theta$, and arrive at \begin{equation}
\frac{d}{dt}\left(\begin{array}{c}
\delta p\\
\delta \theta\end{array}\right)=\left(\begin{array}{cc}
f_{1} & f_{2}\\
g_{1} & g_{2}\end{array}\right)\left(\begin{array}{c}
\delta p\\
\delta\theta\end{array}\right)\equiv\mathcal{M}\left(\begin{array}{c}
\delta p\\
\delta\theta\end{array}\right)\,,\end{equation} where matrix elements of $\mathcal{M}$ are given in the Supplementary Material \cite{sup}. 
 If any of the eigenvalues of $\mathcal{M}$ has a positive real part, the fluctuation terms $\delta p$ and $\delta\theta$ grow exponentially in time and therefore the system is dynamically unstable \cite{dyn}. For unstable states, we denote the largest real part as $\gamma$, which can be regarded as the decay rate of the unstable states.

\begin{figure}[htp]
\includegraphics[width=.48\textwidth]{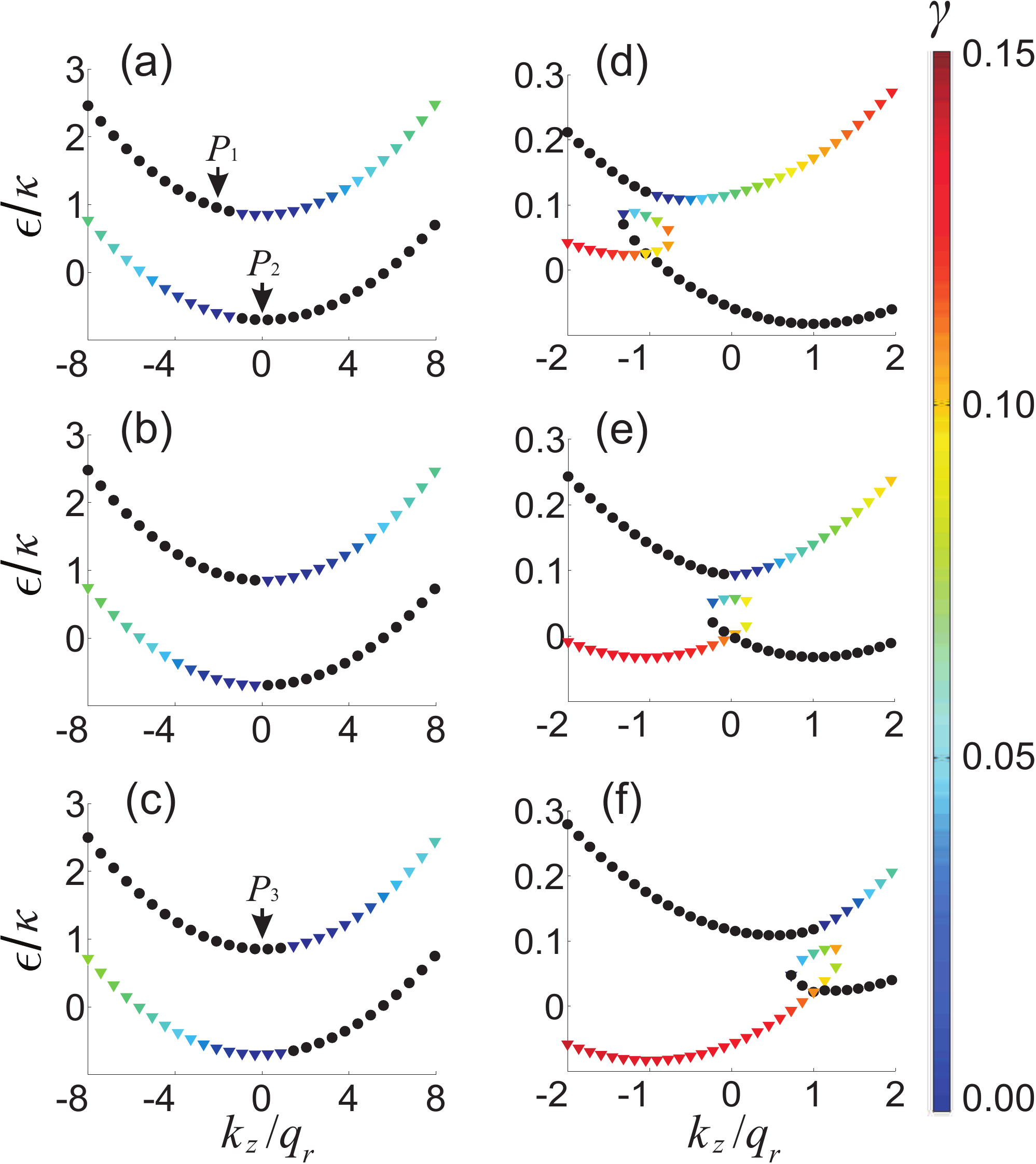}\caption{(Color Online) Stability analysis of the dispersion curve. Colored triangles represent dynamically unstable states and black solid dots represent dynamically stable ones. The colorbar represents  $\gamma$, the decay rate of the unstable states. In all figures, $\Omega=1.1\kappa$ and $\delta_c=\kappa$. From (a) to (c) $\varepsilon_p=2\kappa$ and $\tilde{\delta}=0.05$, 0, and $-0.05\kappa$; from (d) to (f) $\varepsilon_p=0.2\kappa$, and $\tilde{\delta}=0.05$, 0, $-0.05\kappa$. 
}\label{stability}
\end{figure}

A typical result of the stability analysis is shown in Fig.~\ref{stability}, where we plot the dispersion curves and indicate the stability of the states using colored triangles. One can observe: (1) in the regime without the loop, one branch is stable and the other branch is unstable; (2) in the regime with the loop, there may exist one or two stable branches and correspondingly three or two unstable branches. This means that the cavity feedback completely alters the system's stability. Nevertheless, for a relatively large cavity pumping rate as shown in Fig.~\ref{stability}(a)-(c), $\gamma$ is small and the unstable branches are more robust compared with the case represented in Fig.~\ref{stability} (d)-(f) where a smaller $\varepsilon_p$ is used. This can be understood as follows: as the cavity pump rate $\varepsilon_p$ is increased, the cavity photon number increases and the back-action from the atom to the photon becomes less important. Therefore we expect (and have confirmed from our calculation) that in the strong pump limit, the cavity system would not be very different from the conventional system without a cavity \cite{lin, Zhang, mit}.

\begin{figure}[htp]
\includegraphics[width=.48\textwidth]{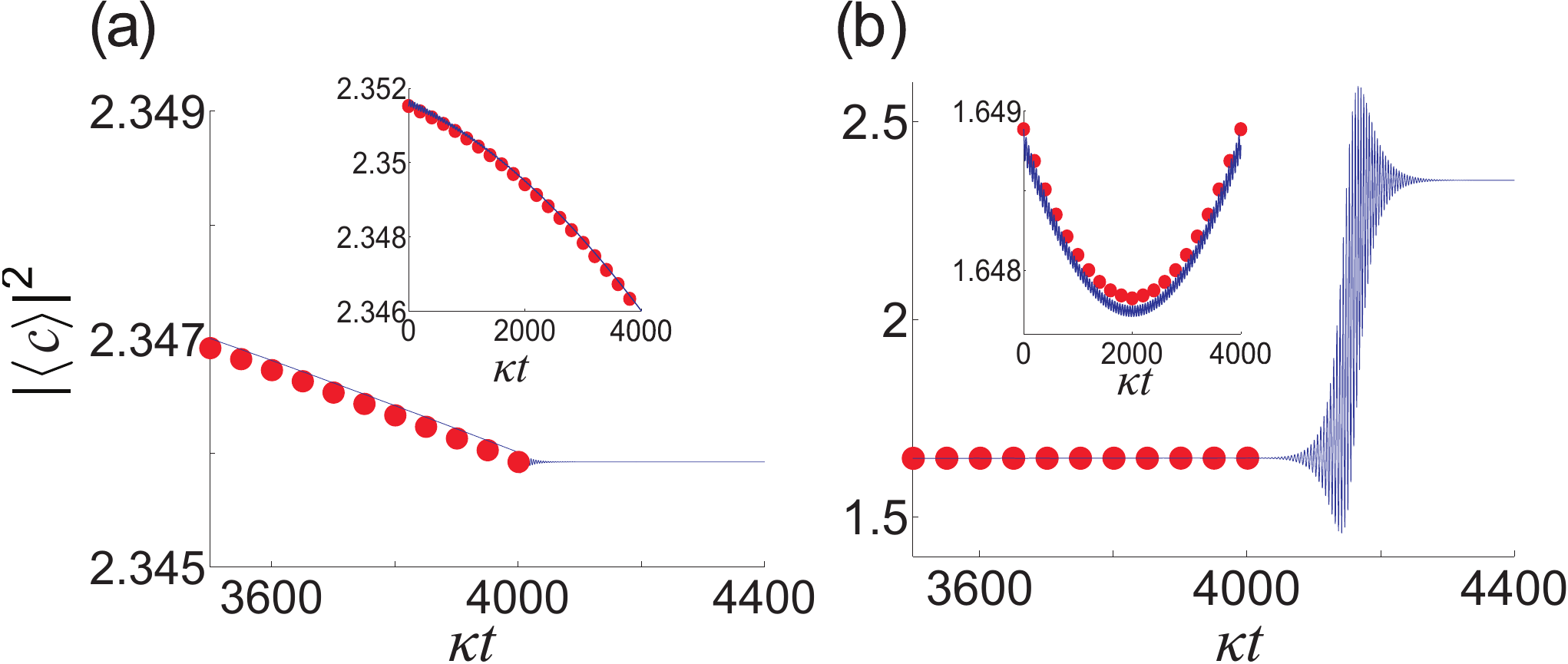}\caption{(Color Online) Evolution of photon number. The initial states are prepared using the same set of parameters as in Fig.~\ref{stability}(a). In (a), we start from point $P_1$ with $k_z=-2q_r$ and in (b) we start from point $P_2$ with $k_z=0$, both indicated in Fig.~\ref{stability}(a). From $t=0$ to $4000/\kappa$, $\tilde{\delta}$ is linearly tuned from $0.05\kappa$ to $-0.05\kappa$ and remains fixed afterwards. Red solid dots represent the photon number corresponding to the instantaneous eigenstate, while blue solid lines represent the dynamical evolution according to 
Eqs.~(\ref{EOMc}) and (\ref{EOMpsi}) after mean-field approximation. 
}\label{photonjump}
\end{figure}

A direct way to detect dynamical instability experimentally in this system is to count the sudden change in the cavity photon number. As an example, we consider the following situation. We start from a stable eigenstate represented in Fig.~\ref{stability}(a). From $t=0$ to $4000 /\kappa$, the two-photon detuning $\tilde{\delta}$ is changed linearly from 0.05$\kappa$ to $-0.05 \kappa$ and remains fixed at $-0.05 \kappa$ afterwards. We plot the evolution of the photon number in Fig.~\ref{photonjump}. In Fig.~\ref{photonjump}(a) we start from the state referred to as $P_1$ in Fig.~\ref{stability}(a). During the whole evolution, the photon number follows the corresponding value of the instantaneous eigenstate as the system remains dynamically stable. In Fig.~\ref{photonjump}(b) we start from the state referred to as $P_2$ in Fig.~\ref{stability}(a). During the linear ramp of $\tilde{\delta}$, the photon number follows the corresponding values of the instantaneous eigenstate. However, at the end of the ramp, the system evolves into a dynamically unstable state. The dynamical instability sets in some time after the end of the ramp and the photon number jumps to a different value after a short transient time. The final state matches the stable state $P_3$ with the same atomic quasi-momentum as indicated in Fig.~\ref{stability}(c) (note that the quasi-momentum does not change during the time evolution). 

\emph{Conclusion and Outlook} --- We have considered a system consisting of a single atom (or a non-interacting condensate) whose two hyperfine spin ground states are Raman coupled by two light fields, one of which is a quantized cavity field. In this setting, the internal and external degrees of freedom of the atom and the cavity field are dynamically coupled. This coupling leads to a dynamic SOC for the atom. We have calculated the atomic dispersion relation and examined its stability and dynamic properties. In comparison to the static SOC generated by two classical laser beams which are not affected by the atomic dynamics, the cavity feedback dramatically modifies the properties of the system. Besides giving rise to new physics in the study of synthetic SOC in cold atoms, our system also represents a new cold-atom-based cavity optomechanical system. From a practical point of view, all the ingredients proposed in this work have been demonstrated in various labs. Hence our proposal can be readily tested in experiment. In fact, recent work reported in Refs.~\cite{cv3,zimm} investigated a BEC inside a ring cavity. A straightforward modification can be used to study the physics predicted in our work. In the future, it will be interesting to extend the study to include inter-atomic interactions and to the case where a system of fermions are Raman coupled via cavity fields. Our current work on the single-particle physics will serve as an important first step towards understanding the many-body properties of the system. It will also be interesting to go beyond the mean-field approximation adopted in our current study \cite{feder}. This is particularly important for very small cavity photon numbers.  

H.P. acknowledges discussions with Su Yi, D. O'Dell and B. P. Venkatesh, and we thank I. White for useful comments on the manuscript. H.P. is supported by the NSF and the Welch Foundation (Grant No. C-1669); B.W. is supported by the NBRP of China (2012CB921300,2013CB921900) and
the NSF of China (11274024, 11334001); L. Z. is supported by the NSF of China (11004057, 11374003), and Shanghai Rising-Star Program under Grant No. 12QA1401000. 

\emph{Note added}: As we were preparing the manuscript, we became aware of the work reported in Ref.~\cite{BECQEDSOC}, where the authors considered cavity-mediated SOC in a transversely pumped standing-wave cavity. Nevertheless, our system is very different from theirs.


\begin{thebibliography}{4}

\bibitem{CQED}  Raimond, J. M., M. Brune, and S. Haroche, Rev. Mod. Phys. {\bf 73}, 565 (2001);  R. Miller, T. E. Northup, K. M. Birnbaum, A. Boca, A. D. Boozer and H. J. Kimble, J. Phys. B: At. Mol. Opt. Phys. {\bf 38}, S551 (2005); H. Walther, B. T. H. Varcoe, B.-G. Englert and T. Becker, Rep. Prog. Phys. {\bf 69}, 1325 (2006).

\bibitem{cv1}F. Brennecke, T. Donner, S. Ritter, T. Bourdel, M. Kohl, and T. Esslenger, Nature, {\bf 450}, 268 (2007).

\bibitem{cv2} Y. Colombe, T. Steinmetz, G. Dubois, F. Linke, D. Hunger, and J. Reichel, Nature {\bf 450}, 272 (2007).

\bibitem{cv3}S. Slama, S. Bux, G. Krenz, C. Zimmermann, and Ph. W. Courteille, Phys. Rev. Lett. {\bf 98}, 053603 (2007).
\bibitem{zimm}D. Schmidt, H. Tomczyk, S. Slama, C. Zimmermann, arXiv:1311.2156.

\bibitem{cv4}S. Gupta, K. L. Moore, K. W. Murch, and D. M. Stamper-Kurn, Phys. Rev. Lett. {\bf 99}, 213601 (2007).

\bibitem{collect2} K. Baumann, C. Guerlin, F. Brennecke, and T. Esslinger, Nature {\bf 464}, 1301 (2010).

\bibitem{dan1}N. Brahms, T. Botter, S. Schreppler, D. W. C. Brooks, and D. M. Stamper-Kurn, Phys. Rev. Lett. {\bf 108}, 133601 (2012); T. Botter, D. W. C. Brooks, S. Schreppler, N. Brahms, and D. M. Stamper-Kurn, Phys. Rev. Lett. {\bf 110}, 153001 (2013).

\bibitem{collect1} Lewenstein, M., A. Sanpera, V. Ahufinger, B. Damski, A. Sen De, and U. Sen,  Adv. Phys. {\bf 56}, 243 (2007).

\bibitem{review}I. B. Mekhov, and H. Ritsch, J. Phys. B: At. Mol. Opt. Phys. {\bf 45}, 102001 (2012).

\bibitem{lin}Y.-J. Lin, K. Jimenez-Garcia, and I. B. Spielman, Nature (London) {\bf 471}, 83 (2011);  Y.-J. Lin, R. L. Compton, K. Jimenez-Garcia, W. D. Phillips, J. V. Porto and I. B. Spielman, Nature Physics, {\bf 7}, 531 (2011).

\bibitem{Zhang} P. Wang, Z.-Q. Yu, Z. Fu, J. Miao, L. Huang, S. Chai, H. Zhai, and J. Zhang, Phys. Rev. Lett. {\bf 109}, 095301 (2012).

\bibitem{mit} L. W. Cheuk, A. T. Sommer, Z. Hadzibabic, T. Yefsah, W. S. Bakr, and M. W. Zwierlein, Phys. Rev. Lett. {\bf 109}, 095302 (2012).

\bibitem{soc_reviews} J. Dalibard, and F. Gerbier, and G. Juzeli\ifmmode \bar{u}\else \={u}\fi{}nas and P. \"Ohberg, Rev. Mod. Phys. {\bf 83} 1523 (2011); N. Goldman, G. Juzeliunas, P. Ohberg, I. B. Spielman, arXiv:1308.6533 (2013).

\bibitem{note} Our scheme is different from a recent proposal \cite{gaugeTheory} where the inter-atomic interaction gives rise to a density-dependent meanfield shift which generates a backaction between the atomic dynamics and the artificial gauge field.  

\bibitem{gaugeTheory} M. J. Edmonds, M. Valiente, G. Juzeliunas, L. Santos, and P. \"{O}hberg, Phys. Rev. Lett. {\bf 110}, 085301 (2013).

\bibitem{sup} See Supplementary Material for details.






\bibitem{photonblock} K. M. Birnbaum, A. Boca, R. Miller, A. D. Boozer, T. E. Northup and H. J. Kimble, Nature {\bf 436}, 87-90 (2005).

\bibitem{note1} For $\Omega<4\epsilon_p$, the solution $\epsilon=0$ corresponds to an unphysical state with $\varphi_\uparrow=\varphi_\downarrow=0$.


\bibitem{loopPapers} B. Wu, and Q. Niu, Phys. Rev. A {\bf 61}, 023402 (2000); E. J. Mueller, Phys. Rev. A {\bf 66}, 063603 (2002); M. Machholm, C. J. Pethick, and H. Smith, Phys. Rev. A {\bf 67}, 053613 (2003); G. Watanabe, S. Yoon, and F. Dalfovo, Phys. Rev. Lett. {\bf 107}, 270404 (2011). 

\bibitem{loop1}J. M. Zhang, W. M. Liu, and D. L. Zhou, Phys. Rev. A {\bf 78}, 043618 (2008); L. Zhou, H. Pu, H. Y. Ling, and W. Zhang, Phys. Rev. Lett. {\bf 103}, 160403 (2009); L. Zhou, H. Pu, H. Y. Ling, Keye Zhang and W. Zhang Phys. Rev. A {\bf 81}, 063641 (2010); Y. Dong, J. Ye, and H. Pu, Phys. Rev. A {\bf 83}, 031608(R) (2011); B. Prasanna Venkatesh, J. Larson, and D. H. J. O'Dell, Phys. Rev. A {\bf 83}, 063606 (2011).


\bibitem{dyn} B. Wu and Qian Niu, Phys. Rev. A {\bf 64}, 061603(R) (2001). 

\bibitem{feder}F. Mivehvar, and D. L. Feder, arXiv:1310.0884 (2013).

\bibitem{BECQEDSOC} Y. Deng, J. Cheng, H. Jing, S. Yi, arXiv:1309.0606v1 (2013). 



\end{thebibliography}
\end{document}